\documentclass[lettersize,journal]{IEEEtran}

\PassOptionsToPackage{hyphens}{url}
\usepackage{setspace}
\usepackage{amsthm}
\usepackage{amssymb}

\usepackage{makecell}
\usepackage{longtable}
\usepackage{booktabs}
\usepackage{float}

\usepackage{bbding}
\usepackage[ruled,vlined]{algorithm2e}
\usepackage{algpseudocode}
\usepackage{amsmath,amsfonts}
\usepackage{array}
\usepackage[caption=false,font=normalsize,labelfont=sf,textfont=sf]{subfig}
\usepackage{textcomp}
\usepackage{stfloats}
\usepackage{url}
\usepackage{verbatim}
\usepackage{graphicx}
\usepackage{cite}
\usepackage{multirow} 
\usepackage[colorlinks,linkcolor=red,anchorcolor=green,citecolor=blue]{hyperref}
\usepackage{cleveref}
\usepackage{flushend}
\usepackage{colortbl,xcolor,pifont} 
\crefname{figure}{fig}{figures}
\Crefname{figure}{Fig}{Figures}
\begin{document}

\title{Toward Edge General Intelligence with Multiple-Large Language Model (Multi-LLM): Architecture, Trust, and Orchestration}

\author{Haoxiang Luo, Yinqiu Liu, Ruichen Zhang, Jiacheng Wang, Gang Sun,~\IEEEmembership{Senior Member,~IEEE}, \\ Dusit Niyato,~\IEEEmembership{Fellow,~IEEE}, Hongfang Yu,~\IEEEmembership{Senior Member,~IEEE}, Zehui Xiong,~\IEEEmembership{Senior Member,~IEEE},\\
Xianbin Wang,~\IEEEmembership{Fellow,~IEEE}, Xuemin Shen,~\IEEEmembership{Fellow,~IEEE} 

\thanks{H. Luo, G. Sun, and H. Yu are with the School of Information and Communication Engineering, University of Electronic Science and Technology of China, Chengdu 611731, China (e-mail: lhx991115@163.com; \{gangsun, yuhf\}@uestc.edu.cn).}
\thanks{Y. Liu, R. Zhang, J. Wang, and D. Niyato are with the College of Computing and Data Science, Nanyang Technological University, Singapore 639798 (e-mail: yinqiu001@e.ntu.edu.sg; \{ruichen.zhang, jiacheng.wang, dniyato\}@ntu.edu.sg). }
\thanks{Z. Xiong is with the School of Electronics, Electrical Engineering and Computer Science (EEECS), Queen's University Belfast, Belfast BT7 1NN, U.K. (z.xiong@qub.ac.uk).}
\thanks{X. Wang is with the Department of Electrical and Computer Engineering, Western University, London N6A 5B9, Canada (e-mail: xianbin.wang@uwo.ca).}
\thanks{X. Shen is with the Department of Electrical and Computer Engineering, University of Waterloo, Waterloo N2L 3G1, Canada (e-mail: sshen@uwaterloo.ca).}
 \thanks{The corresponding author: Gang Sun.}}



\maketitle

\begin{abstract}
Edge computing enables real-time data processing closer to its source, thus improving the latency and performance of edge-enabled AI applications. However, traditional AI models often fall short when dealing with complex, dynamic tasks that require advanced reasoning and multimodal data processing. This survey explores the integration of multi-LLMs (Large Language Models) to address this in edge computing, where multiple specialized LLMs collaborate to enhance task performance and adaptability in resource-constrained environments. We review the transition from conventional edge AI models to single LLM deployment and, ultimately, to multi-LLM systems. The survey discusses enabling technologies such as dynamic orchestration, resource scheduling, and cross-domain knowledge transfer that are key for multi-LLM implementation. A central focus is on trusted multi-LLM systems, ensuring robust decision-making in environments where reliability and privacy are crucial. We also present multimodal multi-LLM architectures, where multiple LLMs specialize in handling different data modalities—such as text, images, and audio—by integrating their outputs for comprehensive analysis.
Finally, we highlight future directions, including improving resource efficiency, trustworthy governance multi-LLM systems, while addressing privacy, trust, and robustness concerns. This survey provides a valuable reference for researchers and practitioners aiming to leverage multi-LLM systems in edge computing applications.
\end{abstract}

\begin{IEEEkeywords}
Large language model (LLM), Multiple LLMs, edge computing, trustworthy LLM system, multimodal LLM.
\end{IEEEkeywords}

\section{Introduction} \label{sec-I}

\subsection{Background}

\IEEEPARstart{E}{dge} computing has emerged as a crucial paradigm to bring intelligent services closer to data sources (sensors, cameras, vehicles, drones) in networks \cite{song2025espd}. By processing data at or near the network edge, computational data exchange-related latency and bandwidth usage can be greatly reduced, thus improving quality-of-service for many time-sensitive intelligent applications such as traffic management, emergency response, and autonomous navigation \cite{chen2025towards}. Traditional edge Artificial Intelligence (AI) systems, however, typically rely on specialized narrow models, each designed for a specific task, e.g., object detection or traffic prediction \cite{deng2020edge}.  While effective for targeted problems, these conventional models lack the flexibility and general reasoning capabilities required to address the increasingly complex and dynamic environments, such as in modern urban and aerial ecosystems.

In contrast, large pre-trained models, especially large language models (LLMs), have demonstrated remarkable human-level proficiency in understanding, generating, and reasoning over natural language and multimodal data \cite{ferrag2025generative}, \cite{liang2024generative}, \cite{guo2025survey}.  Their generalist intelligence enables them to perform a wide variety of tasks without task-specific retraining, making them highly attractive for diverse edge applications where versatility and adaptability are essential \cite{chang2024survey}, \cite{zhao2024generative}, \cite{zhang2024generative0}.  As a result, the integration of LLMs with edge computing leads to the emerging concept of Edge General Intelligence (EGI), where edge nodes, such as Internet of Things (IoT) devices, vehicles, drones, and city infrastructure, gain enhanced context awareness, reasoning capabilities, and multimodal interaction \cite{chen2025towards}, \cite{he2025road}, \cite{zhang2025embodied}.  This evolution enables smarter, more autonomous, and adaptive services directly at the edge, reducing reliance on cloud connectivity and accelerating decision-making processes. However, the deployment of LLMs at the edge introduces unique challenges, including limited on-device compute resources, energy constraints, and high communication overhead for cloud interactions \cite{wang2025edge, zhang2025toward}. To alleviate these difficult problems, a few solutions have been recently developed. Researchers adopted techniques such as model compression \cite{ma2023llm}, pruning, quantization, and knowledge distillation \cite{agrawal2025efficient} to reduce the model size and reasoning complexity to adapt to edge devices. Furthermore, some works have designed energy-saving reasoning, scheduling, and dynamic model invocation strategies \cite{li2025prima}, \cite{li2024tpi} rationally allocating tasks to extend the battery life of the equipment. Furthermore, methods such as edge-local inference enhancement and edge-cloud collaborative computing have been proposed to deal with high communication overhead \cite{tian2024edge}.

Just as LLMs empower edge computing, moving towards EGI, researchers have found that there are still limitations when individual LLMs work. Different LLMs, perhaps with different training data, languages, or specialties, can complement each other \cite{wan2024knowledge}.     Indeed, answers from different LLMs to the same query often vary due to their diverse training corpora and model biases \cite{owens2024multi}.     A single LLM faces difficulty in adapting to heterogeneous contexts (e.g., a traffic management LLM might not handle medical queries) \cite{ding2024easy2hard}.     Additionally, individual LLMs can suffer from outdated knowledge or hallucinations if their training data is limited \cite{wang2025wireless, feng2024don}.  Multi-LLM systems have been proposed as a way to push the frontier of EGI by having multiple LLMs collaborate.        We can combine their strengths and mitigate individual weaknesses by harnessing multiple LLMs as an ensemble or a network \cite{lu2024merge}, \cite{shen2024learning}.     For example, Wang et al. leveraged a pool of LLMs (including GPT-3.5, GPT-4, LLaMA variants, and others) to jointly generate comprehensive elderly care plans, outperforming any single model in coverage of topics \cite{wang2025performance}.     Likewise, researchers have begun exploring multi-LLM systems to simulate “group intelligence,” where LLM-based agents discuss or vote on answers to improve reliability \cite{luo2025weighted}.   It is also regarded as the prerequisite foundation for a multi-agent system \cite{huang2024levels}.  Although these efforts are still in their infancy, they indicate a paradigm shift toward collaborative intelligence at the edge. 

\begin{figure}[!t]
   \centering
   \includegraphics[width=3in]{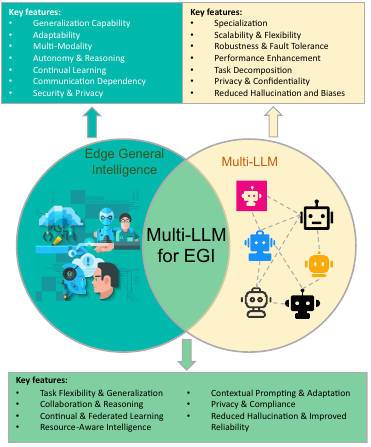}
   \caption{Key features for EGI, multi-LLM, and multi-LLM for EGI.}
   \label{fig0}
\end{figure}

Facing the extensive demand for EGI and recognizing the unique advantages provided by multi-LLM, this paper provides a comprehensive survey of multi-LLM in implementing ubiquitous EGI applications. Before starting the discussion, we summarize the relevant key features of EGI, multi-LLM, and multi-LLM through Fig. \ref{fig0}. The following content will elaborate on this in sequence.

\subsection{Related Surveys}

To clarify the coverage of the existing relevant surveys and highlight our uniqueness, we have summarized the contributions and priorities of the related work. They are divided into two aspects: LLM for EGI and multi-LLM, which are concluded in TABLE \ref{tab1}.

\begin{table*}[!t]
\centering %
    \centering
    \caption{Summary of Related Surveys}
    
    \renewcommand{\arrayrulewidth}{0.8pt} 
    \renewcommand{\tabcolsep}{10pt} 
    
    {\fontsize{8}{10}\selectfont 
     
    \begin{tabular}{m{1.4cm}||m{0.6cm}|m{8.5cm}|m{0.7cm}|m{1cm}|m{1.3cm}} 
        \hline
         \textbf{Scope} & \textbf{Ref.}  & \textbf{Overview} & \textbf{LLM} & \textbf{Multi-LLM} & \textbf{Edge computing}\\ 
        \hline

         \multirow{15}*{LLM for EGI} & \cite{chen2025towards} &   A survey on three conceptual architectures of LLM-licensed EGI: centralized, hybrid, and decentralized, and their implementation methods &\Checkmark&\XSolidBrush&\Checkmark\\ 
        \cline{2-6}
        ~ & \cite{dong2024fine} &  An overview of an efficient memory fine-tuning and model compression in LLM to facilitate its deployment at the network edge &\Checkmark&\XSolidBrush&\Checkmark\\ 
        \cline{2-6}
         ~ & \cite{bhardwaj2024survey} &  A survey emphasizing the role of LLM in reducing latency, enhancing privacy, and improving efficiency for edge computing &\Checkmark&\XSolidBrush&\Checkmark\\ 
        \cline{2-6}
         ~ & \cite{zheng2025review}  & A comprehensive review of edge LLM from resource-efficient model design, pre-deployment strategies, to runtime inference optimization &\Checkmark&\XSolidBrush&\Checkmark\\ 
        \cline{2-6}
         ~ & \cite{wang2025empowering} &  A comprehensive survey on edge LLM deployment technologies, including small language models, model compression, inference optimization, and frameworks &\Checkmark&\XSolidBrush&\Checkmark\\ 
        \cline{2-6}
          ~ & \cite{qu2025mobile} &  A contemporary survey on the structure, caching, delivery, training, and  inference  of deploying LLMs by mobile edge intelligence &\Checkmark&\XSolidBrush&\Checkmark\\ 
        \cline{2-6}
        ~ & \cite{hadish2024language}  & A survey exploring the research trends, developments, and applications of compact edge LLMs &\Checkmark&\XSolidBrush&\Checkmark\\ 
        \cline{2-6}
       ~ & \cite{lin2023pushing} &   A position paper on the deployment of multi-modal LLMs at the 6G edge, including technologies and applications&\Checkmark&\XSolidBrush&\Checkmark\\ 
        \hline
        
      \multirow{8}*{Multi-LLM}  & \cite{lu2024merge} &  A comprehensive survey on different multi-LLM collaboration strategies such as merge, ensemble, and cooperate &\Checkmark&\Checkmark&\XSolidBrush\\ 
        \cline{2-6}
      ~ & \cite{chen2025harnessing} &  A survey on the integration of multiple LLMs before, during, and after reasoning&\Checkmark&\Checkmark&\XSolidBrush\\ 
      \cline{2-6}
      ~ & \cite{feng2025one} &   A position paper emphasizing the multi-LLM collaboration to address the challenges in single LLM, such as reliability, democratization, and diversity&\Checkmark&\Checkmark&\XSolidBrush\\ 
         \cline{2-6}
      ~ & \cite{chen2024role} &  A survey on the relationship (cooperation and competition) between large and small language models&\Checkmark&\Checkmark&\XSolidBrush\\ 
         \cline{2-6}
      ~ & \cite{luo2025trustworthy}&  A conceptual framework on using blockchain to enable multiple LLMs to collaborate and serve wireless networks&\Checkmark&\Checkmark&\XSolidBrush\\ 
        \hline
    \end{tabular}}
    
    \label{tab1}
\end{table*}

\textbf{\emph{1) LLM for EGI:}} Recent literature has increasingly investigated the deployment of LLMs at the network edge to enhance intelligent services with reduced latency and bandwidth consumption. The work in \cite{chen2025towards} presented a comprehensive framework for efficient LLM inference on edge devices by leveraging model compression and hardware acceleration, enabling real-time natural language processing under constrained resources. Dong et al. \cite{dong2024fine} focused on adaptive model pruning and quantization techniques to reduce computational overhead while maintaining inference accuracy, which is crucial for resource-limited edge environments. Furthermore, Bhardwaj et al. \cite{bhardwaj2024survey}  discussed the computing requirements, energy efficiency, and model scalability of deploying LLMs on edge devices with limited resources. Meanwhile, they emphasized the transformative potential and future impact of the combination of LLM and edge computing. In addition, authors proposed a comprehensive overview of the latest advancements in edge LLMs \cite{zheng2025review}. This work covers the entire life cycle of edge LLMs, from the model design and pre-deployment strategies to runtime inference optimization.  The study in \cite{wang2025empowering} explored small language models (SLM), model compression techniques, inference optimization strategies, and dedicated frameworks for edge deployment of LLM. Security and privacy aspects are investigated in \cite{qu2025mobile}, where encrypted model updates and differential privacy mechanisms are introduced for secure LLM deployment on heterogeneous edge devices. Meanwhile, Hadish et al. \cite{hadish2024language} focused on dynamic resource allocation algorithms to adaptively schedule LLM inference tasks across edge clusters, enhancing throughput and energy efficiency. Lastly, Lin et al. \cite{lin2023pushing} systematically surveyed the challenges of LLM deployment at scale on the edge, including hardware heterogeneity, model scalability, and context-aware adaptation, setting a roadmap for future edge intelligence research.

\textbf{\emph{2) Multi-LLM Systems:}} In \cite{lu2024merge}, multiple LLMs interact and cooperate to overcome individual limitations, enabling enhanced reasoning, knowledge sharing, and multi-task capabilities. It also discusses the key challenges and future opportunities in coordinating such multi-model systems. Then, Chen et al. \cite{chen2025harnessing} comprehensively reviewed the emerging field of LLM ensembles, categorizing methods into ensemble-before-inference, ensemble-during-inference, and ensemble-after-inference paradigms.  It analyzed various techniques, related challenges, benchmarks, and applications, highlighting how combining multiple LLMs can improve performance beyond single-model usage. Meanwhile, in the position from \cite{feng2025one}, it argued that a single LLM is insufficient to reliably represent the diverse and complex real-world data, skills, and user populations, advocating instead for multi-LLM collaboration. Moreover, Chen et al. \cite{chen2024role} analyzed the complementary roles of small models alongside LLMs. It highlighted how SLMs contribute to data curation, efficient inference, interpretability, and domain-specific tasks, thus promoting resource-efficient and practical AI deployments. Finally, Luo et al. \cite{luo2025trustworthy} proposed a blockchain-enabled trustworthy multi-LLM network framework to enable collaborative, secure, and reliable responses from multiple LLMs for complex wireless network optimization. It also demonstrated an effective defense mechanism against false base station attacks in 5G or 6G wireless systems.

\begin{figure*}[!t]
   \centering
   \includegraphics[width=5in]{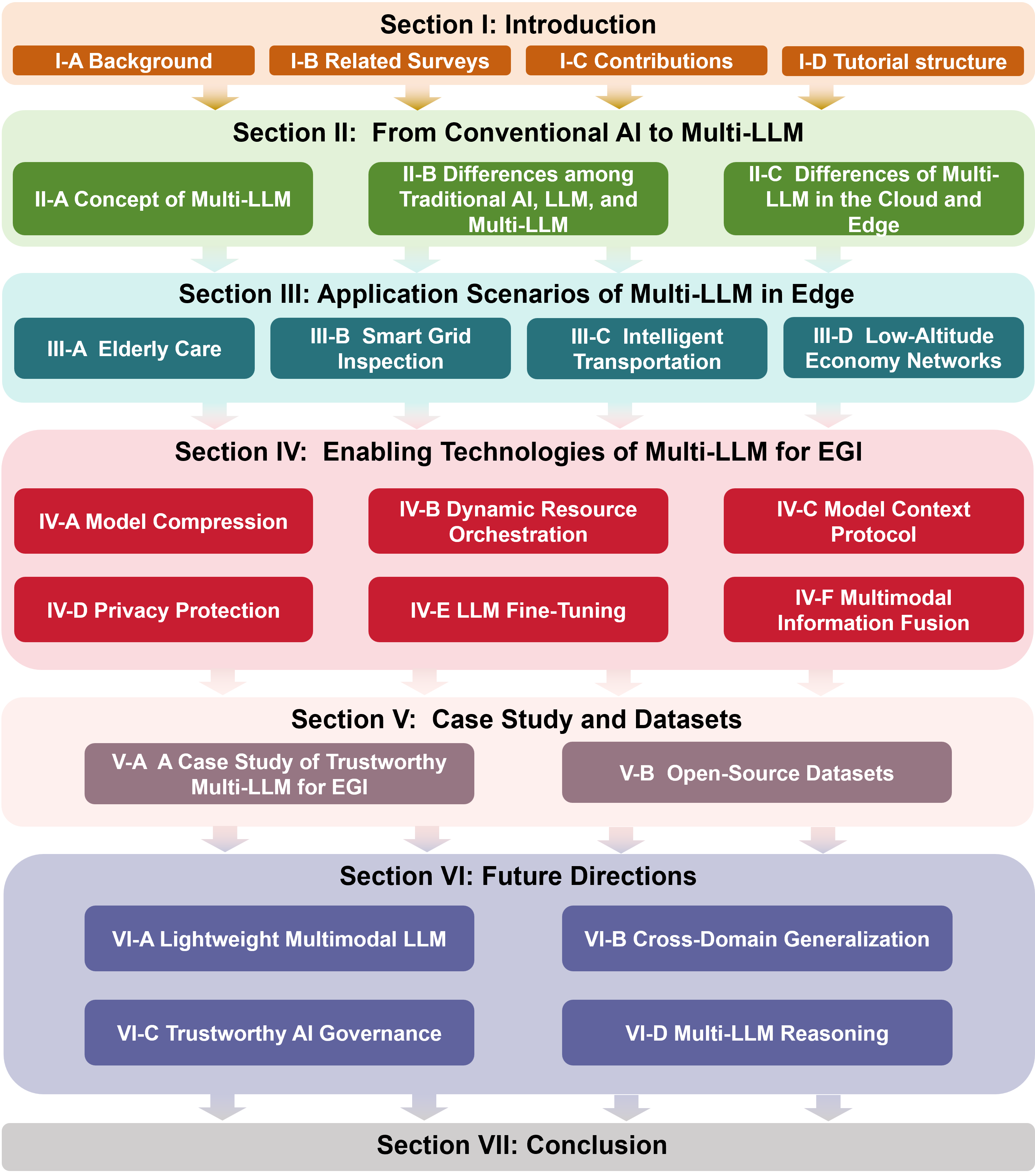}
   \caption{Structure of our survey.}
   \label{fig1}
\end{figure*}

Recent investigations have studied the integration of LLMs in edge computing or the work on the collaboration of multiple LLMs. However, up to now, there has been no comprehensive investigation into the unique opportunities and challenges of multiple LLMs working together in edge networks to achieve EGI. In this survey, our goal is to fill this gap. We have delved into the review of the architecture and technology of multi-LLM. These techniques can achieve powerful and reliable edge intelligence in ubiquitous scenarios.

\subsection{Our Contributions}
The key contributions of this paper are summarized as follows:

\begin{itemize} 

\item We provide a comprehensive review of architectural designs and deployment strategies tailored for multi-LLM systems operating in edge computing environment.      We also investigate the differences among traditional AI, single LLM, and multi-LLM systems.  We comprehensively discuss the differences between running multi-LLM in the cloud and on the edge. Furthermore, we summarize four typical edge applications to reveal the potential of multi-LLM, especially in mobile scenarios such as elderly care, smart grid inspection, intelligent transportation, and Low-Altitude Economic Networks (LAENets).

\item This survey categorizes key enabling technologies based on model compression, resource orchestration, model context protocol, privacy protection, LLM fine-tuning, and multimodal information fusion.      This work lays a foundation for designing a robust multi-LLM-enabled ubiquitous EGI that maximizes local intelligence while reducing reliance on centralized cloud processing.

\item We provide a case study and an open-source dataset summary for deploying multi-LLM systems at the edge. Trustworthiness is a critical factor, where decisions often impact safety-critical or privacy-sensitive applications.  We conduct an in-depth analysis of the mechanisms that ensure reliability and security within multi-LLM collaborations. Then, the available datasets widely used in multi-LLM systems are conducive to promoting further research and experiments in EGI. 



\item Future multi-LLM systems should design lightweight multimodal models that efficiently handle diverse data under edge resource constraints. Therefore, enhancing cross-domain generalization to support knowledge transfer and dynamic adaptation is also essential. Next, establishing a trustworthy AI governance ensures transparency, fairness, securing critical applications. Moreover, the reasoning methods in the multi-LLM system are crucial for further enhancing its ability to handle tasks. Addressing these challenges will enable a strong multi-LLM system, driving advancements in ubiquitous EGI.

\end{itemize}

\subsection{Structure of This Paper}
The structure of this survey is outlined in Fig. \ref{fig1}. Section \ref{sec-II} introduces the concept and design motivation of multi-LLM, including its differences from traditional AI and individual LLM. In Section \ref{sec-III}, it demonstrates four typical applications of multi-LLM. Then, Section \ref{sec-IV} illustrates how edge computing moves towards EGI, that is, the key enabling technologies of multi-LLM empowering EGI. Next, as a tutorial, Section \ref{sec-V}
provides a case study on the trustworthy multi-LLM for EGI and summarizes open-source datasets that support multi-LLM research. Moreover, Section \ref{sec-VI} reveals the future directions of multi-LLM for edge computing. Finally, Section \ref{sec-VII} summarizes this survey.

\section{From Conventional AI to Multi-LLM} \label{sec-II}

This section introduces the multi-LLM concept, including the collaboration methods among LLMs, as well as the differences among traditional AI, individual LLMS, and multi-LLM. The application scenarios of multi-LLM are also discussed.

\subsection{Concept of Multi-LLM}

Multi-LLM systems refer to architectures or frameworks where multiple LLMs operate together (in parallel or in sequence) to solve tasks or generate outputs. In essence, rather than relying on a single monolithic model, a multi-LLM system leverages an ensemble or team of LLMs that work in concert by sharing work or validating each other’s outputs \cite{lu2024merge}. This collaboration allows each model to contribute its specialized strengths toward a shared objective, such as domain knowledge, language, or reasoning skills. 

A key aspect of multi-LLM systems is how the different models collaborate or interact.  Several collaboration modes have been explored in technical literature, each with distinct philosophies for organizing multiple LLMs.  The major paradigms include cooperative approaches, competitive (or adversarial) approaches, and various ensemble-based schemes.  In practice, a multi-LLM system might combine elements of these modes to achieve a balance. Table II summarizes the representative works of these different types of LLM collaborations.  Below, we outline each collaboration mode, clarifying what it entails and providing examples from recent frameworks.

\newcommand{\cooperative}{\textcolor{green}{\scriptsize$\blacksquare$}}    
\newcommand{\competitive}{\textcolor{red}{\scriptsize$\blacklozenge$}}     
\newcommand{\ensemble}{\textcolor{blue}{\scriptsize\ding{108}}}            

\begin{table*}[!t]
\centering
\caption{Comparison of Multi-LLM Works by Collaboration Type}
\label{tab:multi-llm-comparison}
\renewcommand{\arrayrulewidth}{0.8pt} 
    \renewcommand{\tabcolsep}{10pt} 
    {\fontsize{8}{10}\selectfont 
\begin{tabular}{p{0.56cm}| p{8.8cm}| p{5.1cm} |p{0.6cm}}
  \hline
\textbf{Ref.} & \textbf{Contributions} & \textbf{Key Techniques} & \textbf{Type} \\
  \hline

\cite{owens2024multi} & Centralized and decentralized multi-LLM dialogue frameworks reduce bias and improve fairness and accuracy & Multi-model interaction, bias detection, decentralized communication & \cooperative \\
  \hline

\cite{dai2024cost} & Online multi-LLM selection algorithm dynamically balances performance and cost for task scheduling & Multi-armed bandit algorithm, task scheduling, online learning & \cooperative \\
  \hline
\cite{nandkumar2025enhancing}  & Multi-robot navigation via natural language communication, resolving conflicts and prioritizing tasks & Multi-agent dialogue, game theory, control barrier functions & \cooperative \\
  \hline
\cite{yuan2025case} & Multi-LLM consensus with human review framework improves annotation accuracy and reduces human effort & Independent multi-model analysis, consensus mechanism, human-AI collaboration & \cooperative \\
  \hline
\cite{mahadevan2025gamechat}& Multi-level hierarchical LLM conversational agent improves robotic interaction and user satisfaction & Multi-level query classification, hierarchical dialogue management, speech recognition & \cooperative \\
  \hline
\cite{feng2024modular} & Modular pluralism framework uses multi-LLM collaboration for pluralistic alignment & Fine-tuning, token-level multi-LLM interaction, modular decoding strategies & \cooperative \\
 \hline
 \cite{feng2024don} & Multi-LLM competition improves QA accuracy and confidence, effectively reducing hallucinations & Multi-model reflective reasoning, confidence estimation & \competitive \\
  \hline
\cite{chang2024socrasynth} & Socratic multi-LLM dialogue framework enhances reasoning quality through debate and opposing viewpoints & Multi-turn dialogue, debate strategies, reasoning evaluation & \competitive \\
  \hline

  \cite{luo2025weighted} & Blockchain-enabled trusted multi-LLM with weighted Byzantine fault tolerance consensus enhances trust and robustness & Blockchain consensus, decentralized trust evaluation & \ensemble  \\
\hline
\cite{luo2025trustworthy} & Blockchain-enabled trustworthy multi-LLM network mitigates biases and malicious behavior in wireless network optimization & Blockchain consensus, P2P (Peer to Peer) network, multi-LLM cooperative reasoning & \ensemble \\
  \hline
\cite{fang2024multi} & Multi-LLM text summarization framework compares centralized and decentralized strategies, enhancing quality & Centralized evaluation, distributed evaluation, summary generation & \ensemble \\
  \hline
  \cite{mao2025mlsdet} & Multi-LLM statistical deep ensemble framework for high-precision AI-generated Chinese text detection & Mixture of Experts (MoE), statistical feature extraction, cross-entropy metrics & \ensemble \\
  \hline

\cite{fang2025improving} & Sampling-simulation method improves offline multi-LLM inference efficiency with dynamic scheduling and parallelism & Sampling simulation, greedy scheduling, multi-GPU parallelism & \ensemble \\
  \hline
  \multicolumn{4}{l}{\cooperative: cooperative collaboration; \competitive: competitive adversary; \ensemble: ensemble integration} \\
\hline

\end{tabular}}
\end{table*}

\textbf{\emph{1) Cooperative Collaboration:}} Cooperative collaboration describes scenarios where multiple LLMs work synergistically with aligned goals, often dividing complex tasks into subtasks or complementing each other’s strengths. This paradigm leverages role specialization, information sharing, and iterative refinement to enhance overall performance and robustness. 

For instance,  Owens et al. \cite{owens2024multi} investigated centralized and decentralized dialogue frameworks enabling multi-LLM communication to collectively reduce bias and enhance fairness. Dai et al. \cite{dai2024cost} developed an online multi-LLM selection strategy that dynamically allocates queries to different LLMs based on cost-performance trade-offs, facilitating efficient resource utilization in real time. 

In robotics and multi-agent systems, Mahadevan et al. \cite{nandkumar2025enhancing} demonstrated cooperative LLM-based navigation agents that communicate via natural language to resolve conflicts and assign priorities. Yuan et al. \cite{yuan2025case} integrated human reviewers into a multi-LLM consensus pipeline to improve annotation quality while reducing manual effort.  Mahadevan et al. \cite{mahadevan2025gamechat} proposed hierarchical multi-LLM conversational agents that cooperate across different processing levels to improve dialogue efficiency and user experience. Notably,  Feng et al. \cite{feng2024modular} introduced the modular pluralism framework, enabling token-level collaboration between a large black-box LLM and multiple community-tuned LLMs, facilitating pluralistic alignment that better represents diverse societal values.

To summarize, cooperative systems are particularly effective in contexts requiring fairness, bias mitigation, contextual reasoning, and multimodal fusion, providing complementary knowledge and cross-verification capabilities.

\textbf{\emph{2) Competitive  Adversary:}} Not all multi-LLM interactions are purely cooperative. Some frameworks introduce competition or adversarial dynamics between models. In a competitive mode, each LLM prioritizes its objective or hypothesis, which may conflict with others’ objectives \cite{tran2025multi}. Rather than helping each other outright, the models essentially strive to outperform or correct one another, with the system ultimately picking the best outcome from the contest. The rationale is that competition can push each model to perform at its best, yielding a more rigorous solution overall.

A case of competition is the adversarial collaboration approach, where models are explicitly set to challenge each other’s outputs in order to stress-test the correctness and consistency of results \cite{yang2025minimizing}. In adversarial debate frameworks, multiple LLMs serve as “advocates” or debaters for different answers, and they iteratively critique each other’s reasoning \cite{bandi2024adversarial}. This approach essentially transforms the problem-solving process into a dynamic dialogue among LLMs that validates and cross-examines potential answers.

For example, Feng et al. \cite{feng2024don} utilized this approach by enabling multiple LLMs to adversarially critique uncertain responses produced by their peers, mitigating hallucinations and improving answer fidelity. Socratic multi-LLM dialogue \cite{chang2024socrasynth} similarly incorporates adversarial debates, allowing models to refine outputs through dialectical conflict, resulting in enhanced reasoning depth and robustness.

In summary, competitive or adversarial multi-LLM modes introduce a “checks and balances” effect. Each model’s output is vetted by others, promoting robustness at the cost of a more complex interaction protocol.

\textbf{\emph{3) Ensemble Integration:}} Ensemble-based methods are a classical approach adapted to LLMs. Multiple models are run in parallel on the same task, and their outputs are combined to produce a final result. Unlike the interactive cooperation or debate methods above, a pure ensemble often involves little to no direct communication between the LLMs during inference. Each model independently generates an answer (or prediction), and then a predefined aggregation mechanism synthesizes these results \cite{lu2024merge}. The goal is to harness the individual strengths of each model and cancel out their independent errors, improving overall accuracy or reliability.

Typical integration relies on the voting results of each LLM. On the one hand, the average probability distribution of different results can be statistically analyzed from multiple LLMs to determine the final response. For example, Luo et al. \cite{luo2025trustworthy} used voting to decide on the best base station power allocation scheme to resist fake base station attacks for the wireless communication system. On the other hand, there is the weighted voting of confidence level. If the models can provide confidence scores for their answers, the final decision will tend towards the answer with the highest total weight. Luo et al. \cite{luo2025trustworthy} designed a weighted Byzantine fault tolerance to ensure robustness of multi-LLM in wireless network optimization. 

In addition, some other typical examples of integration, including that Fang et al. \cite{fang2024multi} examined ensemble strategies for text summarization, contrasting centralized and decentralized frameworks. Mao et al. \cite{mao2025mlsdet} presented a statistical deep ensemble framework tailored for detecting AI-generated Chinese text with high precision, employing Mixture of Experts (MoE) and statistical feature extraction. Meanwhile, to accelerate the integration of multi-LLM, Fang et al. \cite{fang2025improving} introduced a sampling-simulation method to improve multi-LLM offline inference efficiency, leveraging dynamic scheduling and parallelism across multiple GPUs.              

In general, ensemble methods benefit from the diversity of individual models. Also, it typically yields stable and reliable outputs without tightening model interaction during inference.

\subsection{Differences among Traditional AI, Single LLM, and Multi-LLM}

\begin{figure*}[!t]
   \centering
   \includegraphics[width=7in]{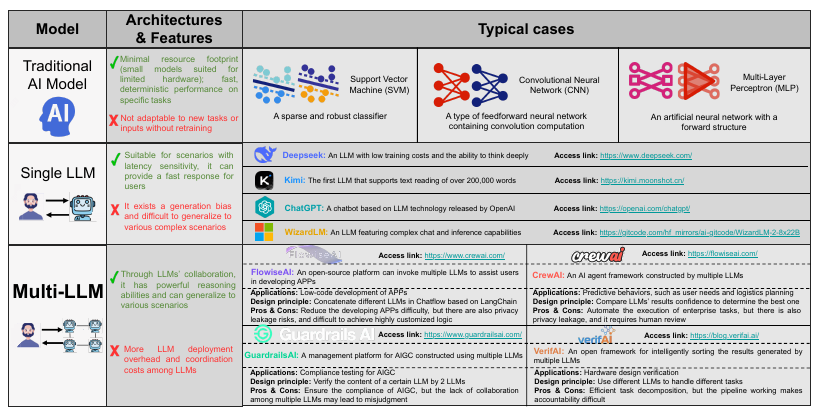}
   \caption{Comparison with traditional AI, single LLM, and multi-LLM system. Traditional AI models require targeted training based on the particularity of scenarios and lack generalization. Due to the limitations of training techniques and data, a single LLM often generates biased and illusory results. Multi-LLM can overcome the above problems through the collaboration of multiple LLMs.}
   \label{fig2}
\end{figure*}

Before applying the multi-LLM system to edge computing, it is necessary to understand its characteristics and advantages. In this part, we compare the differences among traditional AI models, single LLM, and multi-LLM. Fig. \ref{fig2} reveals in detail their characteristics, typical cases, etc.

\noindent \textbf{\emph{1) Traditional AI}} 

Traditional AI models encompass classical machine learning algorithms and compact neural networks that are tailored to specific tasks. These models are typically purpose-built. Each model is trained on a well-defined dataset to perform a particular function, e.g., object detection, anomaly detection, or signal classification. Common designs include decision trees, support vector machines \cite{hsu2025prediction}, and small-scale deep neural networks such as lightweight Convolutional Neural Networks (CNNs) \cite{wong2022energy} or shallow Multi-Layer Perceptrons (MLPs) \cite{ahmed2024secure}. They are selected because they operated efficiently on the limited-resource devices. The underlying design philosophy is to keep models narrow and optimized for the target task, using hand-crafted features or specialized network architectures to maximize accuracy within that domain \cite{curzon2021privacy}, \cite{zha2025data}. In terms of capabilities, traditional models excel at the tasks they are designed for, often achieving high performance under the conditions they are trained on. 

The reasoning and generalization capabilities of these AI models are limited. Because they cannot infer beyond their training distribution, nor do they have the ability to handle tasks beyond their narrow authorization \cite{radha2025reasoning}. For example, a CNN-based traffic classifier can categorize known traffic signs accurately. But it cannot interpret natural language instructions or adapt to recognize new, unrelated objects without retraining.      Each additional functionality (say, adding voice command recognition to a camera system) typically requires developing or deploying a separate model or algorithm.     

Due to their limited scope, traditional AI is usually organized in isolated or pipeline structures rather than as an integrated system \cite{steidl2023pipeline}. Especially at the edge, deployment may involve multiple independent models. Each model handles one aspect of complex tasks. For example, one model is used for face recognition, and another model is used for audio detection. However, any advanced decision logic must be manually encoded or processed by a simple rule-based framework. Therefore, traditional edge artificial intelligence cannot learn dynamically from each other's outputs in an open manner \cite{singh2023edge}.


\noindent \textbf{\emph{2) Single LLM}} 

Single LLM refers to a general-purpose intelligence may based on the Transformer architecture \cite{ma2024megalodon}. Typical representatives include ChatGPT\footnote{https://chatgpt.com/}, WizardLM\footnote{https://github.com/nlpxucan/WizardLM}, Deepseek\footnote{https://www.deepseek.com/}, and Kimi\footnote{https://www.kimi.com/}, etc.  These models are characterized by deep layers of self-attention and very large parameter counts, having been pre-trained on massive text or multi-modal corpora \cite{huang2024large}. Notably, these models learn a broad statistical representation of language and world knowledge during training. Their underlying design leverages unsupervised or self-supervised learning at scale, which endows them with processing capabilities far beyond any single narrow task model \cite{kou2025gia}.   

A single LLM at the edge can serve as a versatile AI agent handling numerous tasks through prompting or few-shot examples, rather than through task-specific reprogramming \cite{zhang2024generative}. Moreover, LLMs demonstrate a degree of general reasoning \cite{wang2025tutorial}. They can solve problems they have not been explicitly trained on by drawing on analogous examples from their training knowledge. For example, an edge-deployed LLM can analyze log data or sensor readings expressed in natural language and provide insights or summaries \cite{zhou2024large}, then switch to answering users’ questions. It is something traditional models cannot do, as LLMs effectively carry a broad prior learned from diverse data. Such cognitive flexibility is highly attractive for edge intelligence scenarios where the types of queries and data can vary widely \cite{zhang2024beyond}.

Deploying a single LLM on edge hardware, however, is challenging due to resource requirements \cite{zhang2024generative1}. These models typically demand gigabytes of memory and significant compute, for example, tens of billions of parameters require specialized hardware accelerators \cite{cao2025toward}. In practice, making an LLM edge-friendly involves model compression techniques like quantization and distillation \cite{xiao2022distill}, reducing precision or size to fit on GPUs or CPUs available at the edge. Recent works introduce SLMs \cite{wang2025empowering}, which are trimmed-down LLMs, on the order of 0.5-7 billion parameters, that aim to retain much of the original’s capability while running on a single edge device. Even so, the inference cost is substantial. The optimized models often still need at least on the order of 500 MB of RAM and an advanced processor to run effectively \cite{cao2025toward}. This can strain devices such as IoT gateways or smartphones, especially under real-time constraints. Edge deployments of LLMs may thus require hardware upgrades, e.g., adding an AI accelerator module or offloading parts of the computation to nearby edge servers \cite{deng2022software}.   Techniques like split inference, that is, partitioning the model layers between device and cloud, are also explored to cope with this demand, albeit introducing complexity in return \cite{patel2024splitwise}.

Despite the remarkable capabilities of single LLMs, their use in edge computing exposes critical limitations. Individual models often produce hallucinations, generating biased or false content, due to incomplete training data or architectural constraints \cite{owens2024multi}, \cite{zhou2024unibias}, \cite{che2025eazy}. These issues not only hinder reliability and adaptability in dynamic network settings but also highlight the single-model approach as a potential bottleneck \cite{zheng2025review}. Consequently, researchers are turning to multi-LLM collaborative systems to overcome these shortcomings.      This paradigm shift, as detailed next, forms a logical step toward more resilient and scalable intelligence at the network edge.

\noindent \textbf{\emph{3) Multi-LLM System}} 

Multi-LLM collaboration involves a network of LLM-based agents working together, potentially offering a form of collective intelligence. Through various types of collaboration methods, namely collaboration, competition, and ensemble, they have avoided the generalization weaknesses, such as illusion and bias of individual LLMs \cite{owens2024multi}, \cite{feng2024don}. This approach finds typical use cases in 360 AI Assistant\footnote{https://bot.360.com} and Corex\footnote{https://link.zhihu.com/?target=https\%3A//github.com/QiushiSun/Corex}. The former is capable of triggering the collaborative operation of three LLMs. Developed by the Shanghai AI Lab, the latter allows multiple LLMs to carry out reasoning in a joint manner. Moreover, Amazon\footnote{https://aws.amazon.com/cn/blogs/machine-learning/multi-llm-routing-strategies-for-generative-ai-applications-on-aws/} has delved into the field of message routing approaches among multiple LLMs. Additional instances are presented in Fig. \ref{fig2}.

The key advantages of the multi-LLM system framework can be summarized as follows:

\begin{itemize} 

\item \textbf{Specialization:} The capabilities of a multi-LLM system arise from the diversity and specialization of its member models. By leveraging different knowledge bases and strengths of each LLM, such a system can cover a broader range of tasks or a more complex task space than any single model. For example, one LLM could be a math expert while another is skilled in understanding legal text. Through collaboration, they can solve problems that involve both legal interpretation and numerical reasoning \cite{stephan2024calculation}.

\item \textbf{Scalability \& Flexibility:} Compared with a single LLM, expansion usually means retraining or fine-tuning the entire model. For multi-LLM systems, new models can be added or replaced without retraining the entire system, thereby adapting to new tasks.

\item \textbf{Robustness \& Fault Tolerance:}  In essence, multi-LLM systems introduce redundancy and specialization.  They are more fault-tolerant and can tackle multifaceted tasks by distributing subtasks among themselves. This approach is inspired by multi-agent system principles, moving AI from isolated models to a collaboration-centric paradigm \cite{tran2025multi}.

\item \textbf{Performance Enhancement:} 
 Even when models are of a similar general-purpose nature, collaboration allows for cross-verification and ensemble effects. An LLM can double-check or critique the output of another, similar to having multiple advisors discuss a problem, which often yields a more accurate and robust result \cite{chang2024socrasynth}.

\item  \textbf{Task Decomposition:} The system’s architecture can take various forms. A centralized scheme might use a leader or supervisor LLM that delegates subtasks to other specialist LLMs. On the contrary, to avoid a single point of failure, a decentralized scheme treats each model as a peer in a distributed protocol, for example, voting or consensus to ensemble results. 

\item  \textbf{Privacy \& Confidentiality:} Through the hierarchical data management of multi-LLM, users' sensitive data can be routed to be stored in a private or secure LLM. While public or non-privacy-sensitive types of queries can use the public model. In a single LLM, all data has to pass through the same LLM, which raises concerns about privacy and compliance. Moreover, LLM can manipulate data without the knowledge of other entities \cite{peng2023generating}.

\item  \textbf{Reduced Hallucination and Biases:} Due to the limitations of a single LLM in terms of training data, technical paths, etc., the generated content may be outdated, biased, and illusory. A system composed of multiple LLMs can effectively avoid this problem, provide users with comprehensive responses, and generalize to various complex scenarios \cite{luo2025weighted}, \cite{luo2025trustworthy}.

\end{itemize} 

It is precisely because of these powerful capabilities, the enhanced reasoning ability, and the more comprehensive information interaction through collaboration among LLMs, that multi-LLM systems are also seen as evolving towards agentic AI \cite{raza2025trism}.
As a result, multi-LLM can bring benefits to EGI, such as task flexibility and generalization, collaborative reasoning, and reduced hallucinations. However, it also introduces many complex challenges. First, the resource footprint grows with each additional model. Running several LLMs in parallel can easily exceed the capacity of the edge side. One solution is to distribute the models across multiple devices, forming an edge cluster \cite{feng2025learning}, but this introduces network communication overhead and requires synchronization. 
Second, coordination mechanisms must be implemented. Unlike a single model, multi-LLM systems need an agreed-upon protocol for collaboration \cite{marro2024scalable}. This might be an internal messaging schema, a turn-taking conversation, or an algorithm for consensus on final answers. Third, there are also trust and consistency challenges. In a decentralized edge setting, how do we trust that each model is providing honest, correct information? Malfunctioning or malicious LLMs could poison the collaborative process. 
As a result, developing and maintaining multi-LLM systems is still an open research area.  

\begin{figure*}[!t]
   \centering
   \includegraphics[width=7in]{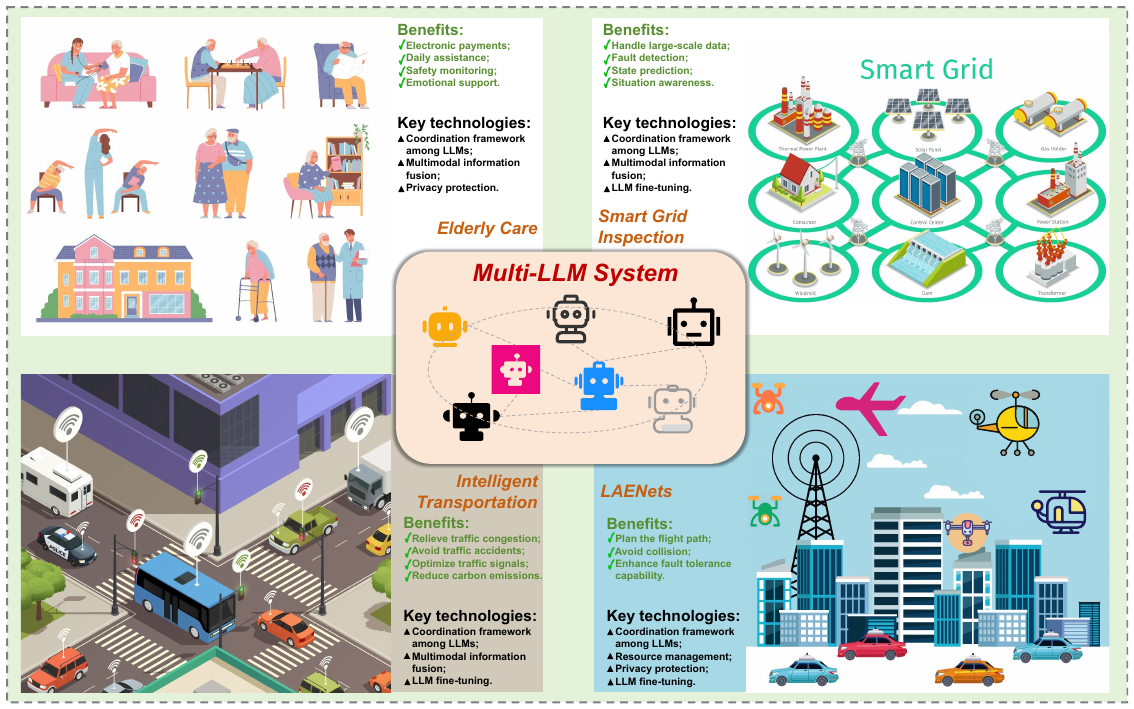}
   \caption{Application scenarios of multi-LLM. It includes typical edge scenarios such as elderly care, smart grid inspection, intelligent transportation, and LAENets.}
   \label{fig3}
\end{figure*}

\subsection{Differences of Multi-LLM in the Cloud and Edge}

Mobile edge deployments of multi-LLM systems differ markedly from cloud deployments across several dimensions. Inference latency is typically lower at the edge because data is processed close to its source, avoiding long network backhauls. By contrast, cloud-based LLM inference incurs additional round-trip delays to distant data centers \cite{zhang2024edgeshard}.

Resource constraints are far more pronounced in edge environments. Edge servers have limited compute, memory, and power, making it challenging to host LLMs.  For instance, a 7 B-parameter LLaMA 2 requires ~28 GB RAM, exceeding most edge devices’ capacity without compression or partitioning \cite{zhang2024edgeshard}. Cloud data centers can leverage abundant GPUs and memory to run such large models more easily.   

Consequently, the system architecture tends to be distributed in edge scenarios. Model inference may be split across multiple edge nodes or performed collaboratively between devices and edge servers \cite{rao2024eco}. In contrast to the centralized architecture in clouds, the entire model ensemble runs in one location. These trade-offs inform applicability. Real-time and privacy-sensitive use cases often favor edge deployments for their immediate responsiveness and local data handling. No need to expose sensitive data to external servers. While cloud-based LLM systems, although they have superior computing capabilities, may have difficulty meeting strict latency needs. 

\section{Application Scenarios of Multi-LLM} \label{sec-III}

To demonstrate the outstanding capabilities of multi-LLM in edge computing, we list four typical applications here, including elderly care, smart grid inspection, intelligent transportation, and LAENets, as shown in Fig. \ref{fig3}. Especially the latter three, reveal the great potential of multi-LLM in mobile scenarios. The extensive application scenarios of multi-LLM have promoted the realization of ubiquitous EGI.

\subsection{Elderly Care}

The multi-LLM system transforms elderly care by serving as smart assistants that address diverse needs in health monitoring, daily living, and social interaction. This application demands multiple LLMs instead of a single model because the tasks span numerous domains (medical advice, emergency alerts, household management, etc.) and modalities (text, speech, sensor data). A single model would be limited in expertise and may have outdated or biased knowledge. While a LLMs team can cover broader knowledge and cross-verify each other’s outputs. For example, one recent system combined GPT-3, GPT-4, and six other LLMs to collectively provide personalized elder care services, handling electronic payments, daily assistance, safety monitoring, and emotional support \cite{wang2025performance}. Such multi-LLM collaboration brings clear benefits. Each LLM can be fine-tuned for a specific function, e.g., a medical domain LLM for health queries, a home automation LLM for smart appliances, a conversational LLM for companionship. By assigning distinct roles to different LLM-based agents and establishing feedback loops, the system can catch errors or hallucinations and improve response quality \cite{ferrimulti}.         

Edge deployment characteristics are crucial in this scenario to ensure responsiveness and trust.       Elderly care assistants are often deployed on distributed edge devices in smart homes or wearable monitors, which keeps sensitive personal and medical data local \cite{syu2023comprehensive}.       On-device processing with multiple cooperating LLMs avoids sending raw data to the cloud, protecting privacy while reducing latency for time-critical support \cite{rahman2021internet}, such as fall detection or medication reminders. The multiple LLMs communicate over the local network or via an edge gateway, requiring efficient coordination protocols to share insights without overwhelming bandwidth. The key supporting technologies include the agent coordination framework, which allows these LLMS to negotiate tasks and share intermediate results. Furthermore, multimodal information fusion requires the integration of camera feeds, biological signals, and natural language inputs to obtain an overall view of the user's state \cite{shaik2024survey}. Equally important is privacy-preserving collaboration \cite{shen2022privacy}.  Techniques such as on-device inference and access control ensure that even as models collaborate \cite{xu2023llmcad}, the user’s personal data remains protected. In this way, collaborative LLMs can provide consistent, accurate, and user-adaptive assistance to the elderly in a trustworthy manner.

\subsection{Smart Grid Inspection}

Smart grid inspection tasks benefit greatly from a multi-LLM approach, including monitoring IoT sensor streams and analyzing drone imagery of power lines \cite{zaboli2024chatgpt}, \cite{shi2024review}. Unlike siloed single-purpose tools, multiple LLMs working in concert can holistically reason over heterogeneous data sources. One LLM alone might be overwhelmed by the volume and variety of grid data or miss cross-modal correlations. By contrast, specialized LLM agents can each focus on a particular modality or sub-problem and then share their findings. For instance, an LLM could correlate a series of voltage sensor anomalies with recent weather conditions, while another examines drone photos of a transformer for physical damage.    Together, they provide a more complete situational picture and can suggest targeted maintenance actions \cite{madani2025large}.   Multi-LLM cooperation yields superior situational awareness for grid operators by cross-verifying insights, which reduces false alarms and improves the accuracy of fault detection and prediction.

Edge deployment is essential for this scenario, given the geographic distribution of the grid and the need for real-time response. Placing LLMs on distributed edge, at substations, control centers, or on inspection drones, minimizes communication delays. It can ensure critical functions can continue locally if cloud connectivity is lost \cite{zhou2022joint}. This arrangement significantly cuts fault response times and reduces bandwidth usage. LLMs communicate over reliable networks or dedicated 5G links, to enhance network security and resilience \cite{carrillo2022boosting}. Key enabling technologies include multimodal data fusion, which combines sensor measurements, weather information and visual inspections into a unified situational awareness. In addition, LLM specialization empowers knowledge of power systems in fine-tuning models \cite{ling2023domain}, enabling them to understand specific grid terms and failure modes. Also, proxy coordination protocols for distributed decision-making are required \cite{ali2022universal}. For example, the edge LLM of the substation can request confirmation of the LLM of the unmanned aerial vehicles (UAVs) before triggering the alarm. Furthermore, achieving real-time performance requires model optimization and dynamic resource scheduling. Heavy computing can be offloaded to more powerful nearby servers. Through the above optimization, the multi-LLM method can immediately perform grid anomaly detection and carry out dynamic grid reconfiguration at the edge, thereby pointing to a more resilient grid.

\subsection{ Intelligent Transportation}

Multi-LLM deployments offer a powerful paradigm for intelligent transportation systems by enabling autonomous traffic management and connected vehicles to collaborate in real time. Urban mobility involves many distributed actors, such as vehicles, traffic signals, and control centers. In highly dynamic conditions, fixed rule-based systems find it difficult to efficiently coordinate these participants. Additionally,  a single centralized AI has limited adaptability and scope, whereas a team of LLMs can divide the cognitive load and respond more flexibly to evolving traffic situations \cite{xu2024genai}. For example, each smart intersection could be managed by its own LLM optimizing local signal timing. A higher-level coordinator LLM oversees city-wide patterns and mitigates congestion by rerouting flows. Vehicles’ LLM coordinates maneuvers with each other and the infrastructure.   This multi-LLM collaboration can reduce congestion and accidents, lower emissions, and provide natural language interfaces for human operators and users. Unlike rigid traditional Intelligent Transportation System (ITS) control, LLM can handle novel scenarios, e.g., unexpected road closures, rather than being limited to pre-programmed responses \cite{mahmud2025integrating}.  

Edge computing is pivotal in this scenario to meet the ultra-low latency requirements of transportation \cite{quan2024federated}. Safety-critical functions, such as collision avoidance or emergency braking coordination, demand millisecond response times. Thus, computation must occur close to data sources, on vehicles and Road Side units (RSUs), or nearby edge servers \cite{jiang2024potential}. This distributed architecture ensures each vehicle or intersection makes instant local decisions while still coordinating with other LLMs for cooperative maneuvers \cite{luo2023esia}. High-speed V2X (vehicle-to-everything) communication enables these LLM agents to exchange information with minimal latency. The key supporting technologies include hierarchical coordination. Some LLMs assume supervisory roles to coordinate the actions of many agents and prevent conflicts \cite{xu2024multi}. Moreover, multimodal perception integration requires the combination of visual sensor data, maps, and other inputs with LLM inference to better understand the traffic environment \cite{chen2025embodied}. LLM specialization can fine-tune some models to accomplish expert tasks such as route optimization, event analysis, or traveler communication. Multi-LLM can deliver smart mobility services, automate tasks like traffic analysis and simulation. By collaborating through both data-driven and natural language exchanges, distributed LLMs collectively make transportation networks safer, more efficient, and more user-friendly.

\subsection{ Low-Altitude Economy Network}

LAENets are emerging as ecosystems of aerial platforms, including delivery drones, urban air mobility vehicles, and UAV-based wireless service nodes \cite{cai2025secure}. They operate in the lower airspace to provide logistics, communication, and sensing services \cite{wang2025toward}.  Managing such a complex network of autonomous agents requires Multi-LLM collaboration to handle the breadth of tasks and the highly dynamic environment. A traditional centralized control or single AI model would be hard-pressed to adapt to constantly changing conditions and uncertainties. In contrast, a multi-LLM-powered distributed team can divide responsibilities and operate concurrently. For example, each drone’s LLM plans its route on the fly;  a network-management LLM optimizes communication links; and a coordinator LLM oversees mission-level objectives while deconflicting routes \cite{zhao2025generative}. This division of labor allows the system to scale and remain robust.  Each LLM makes local decisions using specialized domain knowledge, while sharing essential information such as sensing, navigation, communications, delivery, etc. \cite{cai2025secure}, to collectively adapt as conditions change.  Multi-LLM cooperation improves safety, negotiating the right-of-way to prevent mid-air collisions. It also increases efficiency, collectively scheduling routes to save energy and time \cite{zhu2024task}. Meanwhile, it can boost resilience by avoiding single points of failure.  For instance, if one LLM encounters a problem, others can assist or reroute tasks, providing fault tolerance beyond any monolithic controller \cite{cai2025large}.

Edge computing and communication infrastructure are fundamental to deploying multi-LLM systems in LAENets \cite{mcenroe2022survey}, \cite{luo2024escm}. Each drone is an edge node with limited compute and battery, so only lightweight LLM models run onboard for immediate decisions. Additionally, each drone must act autonomously if a link drops, so local intelligence is indispensable. Key enablers include coordination protocols for LLM-driven drones to share state and plans, such as navigation, communications management, and sensing. Then, security is also critical. Robust authentication can prevent malicious or faulty drones from disrupting the network \cite{luo2024split}. Finally, intelligent resource management allocates communication bandwidth, computation, and energy among UAVs, dynamically scheduling tasks and data flows to maximize network performance \cite{zhao2025temporal}. By deploying collaborative LLMs at the edge, LAENets become far more adaptive and reliable in supporting next-generation drone delivery and aerial network services.

\subsection{Lessons Learned}

The application of multi-LLM in mobile edge scenarios shows great potential. However, based on the analysis of the above four scenarios, there are also many challenges, such as resource constraints and high real-time requirements. With the help of key technologies such as model compression, dynamic resource orchestration, model context collaboration, privacy protection, model fine-tuning, and multimodal information fusion, the efficient deployment of multi-LLM in mobile edge scenarios can be achieved, promoting the ubiquitous EGI.

\section{ Enabling Technologies of Multi-LLM for Edge General Intelligence} \label{sec-IV}

According to the lightweight characteristics for LLM to empower EGI, we classify the key enabling technologies into six categories. The following will elaborate respectively.

\subsection{Model Compression}

\begin{figure*}[!t]
   \centering
   \includegraphics[width=7in]{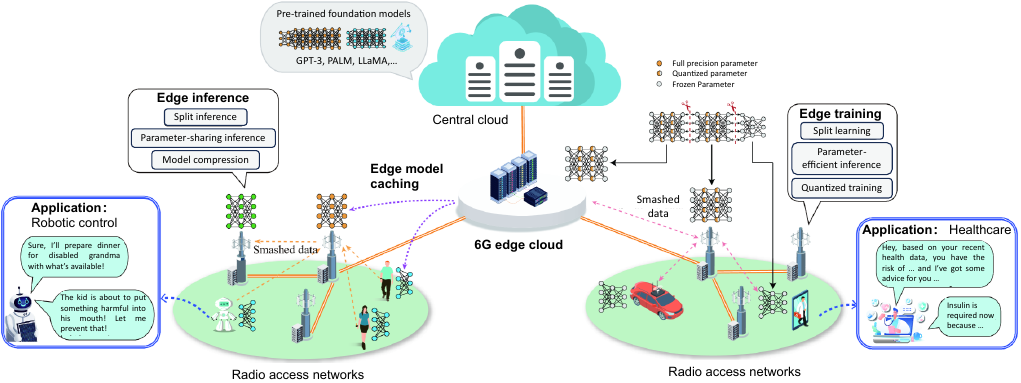}
   \caption{LLM integrating split training in 6G Mobile Edge Computing (MEC) in \cite{lin2023pushing}. This adapter co-training approach keeps user data on-device and drastically cuts communication, since only small adapter weights or token vectors are exchanged.}
   \label{fig4}
\end{figure*}

Edge deployment of LLMs is challenging due to limited device resources, motivating research on lightweight architectures and model compression. Knowledge distillation is a prevalent approach. For example, DistilBERT \cite{sanh2019distilbert} and TinyBERT \cite{jiao2019tinybert} are compact models distilled from BERT (Bidirectional Encoder Representations from Transformers) that preserve most of its accuracy. The former retains 97\% of BERT’s performance with 40\% fewer parameters, and the latter achieves 96.8\% of BERT-base accuracy while being 7.5× smaller and 9.4× faster.  Another strategy is architectural optimization. MobileBERT \cite{hussain2025low} introduces bottleneck structures and a custom intermediate teacher model, yielding a 4.3× smaller and 5.5× faster variant of BERT that matches its accuracy on many tasks. Similarly, ALBERT \cite{lan2019albert} reduces model size by factorizing embeddings and sharing parameters across layers, achieving an order-of-magnitude parameter reduction.   Specifically, ALBERT-large has 18 M vs 334 M parameters in BERT-large with minimal loss in accuracy.  Other techniques include pruning redundant weights and quantizing model weights to reduce the memory footprint and inference latency.  EdgeBERT \cite{tambe2021edgebert} exemplifies a combination of these.   It uses adaptive attention, selective pruning, and quantization to fit BERT on edge devices, yielding up to 7× energy savings in multi-task NLP (Natural Language Processing) inference. Collectively, these innovations drastically reduce LLM model size and computation, enabling LLM deployment within the strict latency and memory constraints of edge computing.     

Beyond single-model compression, lightweight multi-LLM architectures at the edge often coordinate multiple small models to meet real-time and resource constraints. The typical scheme is as follows:

\begin{itemize} 
 \item \textbf{Query routing \& Cascaded Inference:} It can send each request first to a compact model and escalate only hard cases to a larger model \cite{zellinger2025cost}.       Additionally, confidence-based gating or ensemble voting among local LLMs enables distributed decision-making. For instance, in intelligent transportation, an RSU runs a small LLM, and if its prediction confidence is low, the query is forwarded to a more powerful LLM (possibly in the cloud).       Such cascades dramatically reduce average latency and energy by keeping most computation on low-cost models.       In real-world deployments, these ideas appear in traffic forecasting and image-generation systems.       For example, the LSGLLM-E architecture partitions a city’s road network into subregions and deploys a lightweight spatio-temporal generative LLM on each RSU \cite{rong2024large}.    This edge-based LLM captures local spatio-temporal correlations and offloads work from the central cloud, achieving superior accuracy and efficiency.       Similarly, a synergetic big-little framework co-trains a large cloud LLM with multiple small edge LLMs  \cite{tian2024edge}.  The cloud model provides high-level guidance while edge models capture local data, and a distributed training scheme aligns their parameters.

 \item \textbf{Cloud–Edge Co-Training:} On the training side, this scheme uses model-splitting and federated updates to share learning between devices. For example, split learning (SL) \cite{lin2024split} can partition an LLM between edge and server so that only intermediate representations are transmitted. Combining SL with LoRA (Low-Rank Adaptation)-style adapters, multiple edge servers can jointly fine-tune a shared model in parallel. One study shows that integrating LoRA with parallel split federated learning allows large-model fine-tuning on edge GPUs in a reasonable time \cite{lin2023pushing}. Fig. \ref{fig4} shows the technical route of it when applied to the 6G edge.

 \item  \textbf{Lightweight Inference:}  At inference time, lightweight strategies like cascades, adapters, and token filtering further serve edge constraints.  Confidence-based early exits give an edge to LLM output easy tokens immediately and offload only uncertain tokens for cloud processing \cite{jin2024collm}. Modular adapters enable a small model to be quickly customized to local context without invoking the full LLM. Even simple token pruning or split inference can cut latency. Placing the transformer encoders at the edge and only sending compact token embeddings to the cloud means that only tokens need to be exchanged.
 
\end{itemize} 
 
Together, these techniques meet stringent edge requirements.   They yield low-latency answers and reduced energy and communication costs by keeping most work on-device.

\subsection{Dynamic Resource Orchestration}

Edge-based LLM systems often employ early-exit and adaptive resource strategies to meet strict latency/energy targets on constrained devices \cite{zhang2025deadline}.  For example, EdgeBERT uses an entropy-based early-exit policy to terminate inference at intermediate layers \cite{tambe2021edgebert}. When confidence is high, it can significantly cut computation and latency. Similarly, CE-CoLLM \cite{jin2024collm} adopts a two-mode, cloud-edge pipeline.   In a low-latency edge-only mode, it runs a lightweight early-exit LLM locally. While in a collaborative mode, high-confidence tokens are handled on-device and only low-confidence tokens (or residual computation) are offloaded to a larger cloud model. These approaches exemplify single-LLM orchestration.  By combining early exits with adaptive scheduling and judicious model splitting between edge and cloud, they meet real-time constraints while saving energy and bandwidth \cite{zheng2025review}, \cite{li2021slicing}.

Distributed LLM inference extends these ideas across multiple devices or model replicas using advanced routing and scheduling policies. 

\begin{figure*}[!t]
   \centering
   \includegraphics[width=7in]{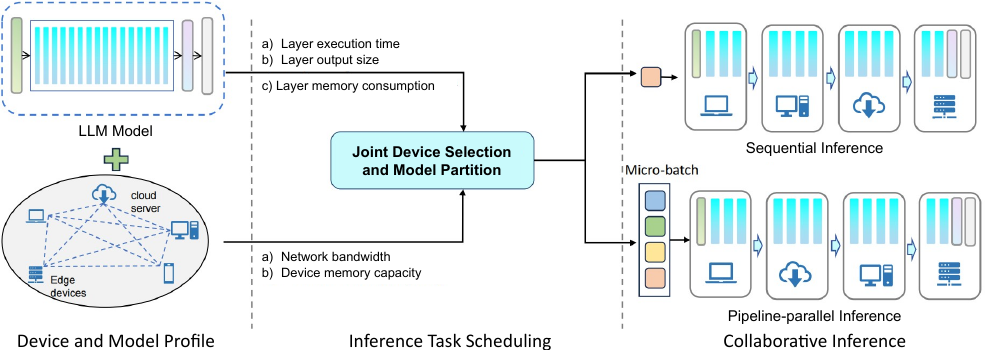}
   \caption{The framework of EdgeLLM in \cite{zhang2024edgeshard}. It includes offline analysis, task scheduling optimization, and online collaborative LLM inference.}
   \label{fig5}
\end{figure*}

\begin{itemize} 
 \item \textbf{Learned routing Policy:} It assigns queries to different models based on difficulty. For instance, Hybrid LLM \cite{ding2024hybrid} trains a router to steer queries to a small or large model by predicting each query’s hardness, dynamically trading off cost and accuracy. In a similar spirit, Agreement-Based Cascading (ABC) builds a cascade of increasingly powerful models and uses ensemble agreement at each stage \cite{kolawole2407agreement}. If a small-model ensemble agrees confidently, inference stops locally. Otherwise, it cascades the query to a bigger model. ABC’s data-dependent routing yields large savings in cloud traffic while preserving accuracy.

  \item \textbf{Edge-oriented Framework:}  It also performs cooperative scheduling and load balancing across devices. EdgeShard, for example, formulates a joint device-selection and partition problem. It shards the LLM across heterogeneous edge nodes and uses a dynamic-programming algorithm to allocate layers to devices, so as to minimize latency and maximize throughput \cite{zhang2024edgeshard}, as shown in Fig. \ref{fig5}.               Likewise, Tian et al.’s synergetic big-cloud/small-edge architecture \cite{tian2024edge} trains and deploys a large cloud model alongside many tiny edge models. It uses a distributed task-oriented protocol, so that lightweight edge models handle common tasks locally.        While the big model refines or assists only when needed, it harnesses collaborative intelligence to reduce latency and network load. 
 
  \item \textbf{Multi-Agent Scheduling :} The MASITO framework \cite{ben2024multi} uses cooperating DRL (Deep Reinforcement Learning) agents on each edge server to schedule inference tasks and offloading under time/energy constraints.      Then, it can effectively balance load and optimize accuracy across the network.

 \end{itemize}

In summary, edge LLM orchestration now spans learned routers, cascaded ensembles, cooperative load distribution, and multi-agent DRL-based task allocation.  All are designed balancing efficiency, accuracy, and latency in heterogeneous edge environments.

\subsection{Model Context Protocol}

The information transmission between LLMS and between LLMS and user cannot be done without the Model Context Protocol (MCP) \cite{ray2025survey}. Recent work on LLM context management focuses on compressing or streaming the input to fit edge constraints. For example, the In-Context Autoencoder (ICAE) trains a light-weight autoencoder to summarize a long input context into a small set of memory slots. It achieves roughly 4× context compression with minimal extra parameters \cite{ge2023context}. This allows the LLM to condition on a much shorter representation of the full history, reducing GPU memory and latency. Similarly, EdgeInfinite \cite{chen2025edgeinfinite} introduces a trainable memory-gating module within the Transformer. During inference, it keeps a few attention sink tokens and a sliding window of recent tokens in the KV (Key Value) cache, while compressing older tokens into a compact memory block. This gating mechanism is pretrained to preserve key semantic and positional information, enabling effective infinite-context inference on edge devices with little overhead. 

\begin{figure}[!t]
   \centering
   \includegraphics[width=3in]{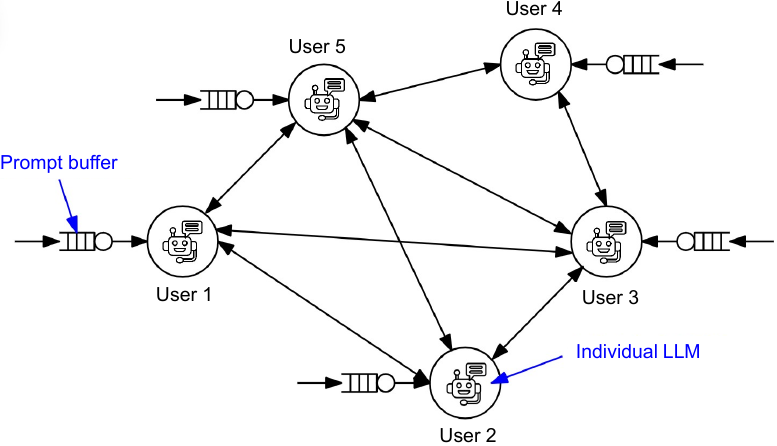}
   \caption{Distributed MoA system model in \cite{mitra2024distributed}. Each device has its local prompt and makes inferences through the local LLM. Meanwhile, these prompts are sent to neighboring LLMs for reasoning, and then their responses are aggregated by one LLM.}
   \label{fig6}
\end{figure}

Coordinating multiple LLMs on the edge introduces additional protocol complexity for sharing context among models. The specific methods are:

\begin{itemize}
    \item \textbf{Model-Distributed Inference:} One strategy is model-distributed inference, exemplified by MDI-LLM \cite{macario2025model}. In this approach, a single generation task is split across devices. The LLM’s layers are partitioned among nodes, which pipeline their computation to reduce idle time. 

    \item \textbf{Separate Private and Public Memory Banks:}  Another class of protocols treats memory by separating private vs. shared memory banks.        For instance, the Collaborative Memory framework \cite{rezazadeh2025collaborative} proposes a two-tier memory architecture per LLM.        Each model instance on an edge device maintains a private memory for its user-specific interactions and a shared memory for knowledge that can benefit others. When an LLM completes a query, it decides whether to write the interaction into shared memory or keep it private.        Under limited edge resources, this selective sharing prevents unnecessary duplication while still enabling cross-LLM context reuse.

    \item \textbf{Retrieval-based Memory Fusion:}  It is another related technique. For example, in the Memory Sharing (MS) framework \cite{gao2024memory}, each agent logs its queries and answers as discrete memories.        A communal memory store aggregates (prompt, response) pairs from many LLMs.        When an LLM answers a new prompt, it performs retrieval over this shared pool to find relevant examples from other agents’ experiences.        These retrieved examples are then concatenated into the in-context prompt, effectively fusing external context into the local inference.        In \cite{gao2024memory}, MS aids agents in identifying the most relevant examples for specific tasks by pooling memories across the network.

    \item \textbf{P2P Prompt Exchange Protocol:}  This type of protocol represents the exchange of context among LLMs without a central server \cite{luo2025weighted, luo2025trustworthy}. In the Mixture-of-Agents (MoA) system \cite{mitra2024distributed}, each device runs its LLM, and the devices intermittently communicate, as shown in Fig. \ref{fig6}. When a user generates a query, the device may broadcast that prompt to neighboring LLMs.   Meanwhile, one LLM collects these multiple proposals and resolves them into a final answer.       The P2P network ensures that updates propagate locally, but edge nodes have limited memory, the system must limit how many prompts sit in queues.           
\end{itemize}

\subsection{Privacy Protection}

In edge deployments of a single LLM, privacy protection often relies on minimizing data exposure and securing on-device computation.          For example, data sanitization can strip personal identifiers from user inputs.  As demonstrated by Rescriber \cite{zhou2025rescriber}, a system that helps users redact personal data from prompts and thereby reduces unnecessary disclosure.          Local inference is also a key measure.       By running the model entirely on the edge device, no raw user data needs to be sent to external servers, thereby keeping all personal inputs localized \cite{yan2024protecting}.          Additionally, advanced hardware and cryptographic safeguards are often employed.      For instance, Trusted Execution Environments (TEE) encrypt and isolate model execution to protect data in use \cite{li2024teeslice}.          Finally, Differential Privacy (DP) techniques can be applied to model outputs or intermediate representations. Injecting calibrated noise into on-device embeddings can prevent user text from being exactly reconstructed downstream \cite{mai2023split}.   Together, these measures, including data minimization, local execution, encryption, and DP, form a layered defense that has become standard in edge LLM inference.

In edge-based multi-agent or multi-model settings, enhanced privacy measures are essential. 

\begin{itemize}
    \item \textbf{Federated Learning:} It has been widely proposed for LLMs to train or fine-tune shared models without exchanging raw data \cite{wu2024fedbiot}, \cite{zhang2021optimizing}.   Each device updates a local copy of the LLM, and only encrypted model deltas or gradients are shared, keeping user data on-device.   However, even gradient updates can inadvertently leak information, so federated LLM frameworks typically layer on extra defenses.   For instance, differential privacy or cryptographic masking can be applied to updates. They may be randomly perturbed or aggregated through secure multi-party computation so that individual contributions remain obscure \cite{li2024teeslice}.

    \item \textbf{Cryptographic Methods:}  Likewise, this type of method allows LLMs to jointly infer without revealing inputs.            One example is secure multi-party decoding, which confines user prompts to a TEE when collaborating with another LLM \cite{gim2024confidential}.            Emerging zero-knowledge proof techniques have also been explored to enable verification of LLM outputs without disclosing the inputs \cite{watanabe2025generating}.      
      \item \textbf{Role-based or Context-Aware Policies:}
     These policies can restrict cross-LLM data flow.  For example, Shi et al. \cite{shi2025privacy} introduced the Embedded Privacy-Enhancing Agents (EPEAgents).  In this framework, each LLM might declare its role, and an embedded privacy LLM can filter communications so that only context-relevant information is shared with each model.

\end{itemize}


 Furthermore, LLMs notoriously memorize training examples. Thus, adversarial prompts can trigger a model to regurgitate training data, undermining privacy protections \cite{li2024teeslice}. These cross-LLM leakage and adversarial memorization attacks demonstrate that federated or encrypted LLMs still need robust DP, secure aggregation, and strict access controls to fully protect user data in multi-LLM edge deployments.

\subsection{LLM Fine-Tuning}

Edge computing provides a decentralized computing paradigm, bringing LLMs closer to data sources and users. In general, LLM fine-tuning for edge computing, the following aspects are mainly focused on. First, LoRA freezes the pre-trained weights of the LLM and introduces trainable low-rank matrices to adapt the model to specific edge tasks. For example, in some smart home applications, LoRA can be used to fine-tune LLMs on edge devices to better understand and respond to user voice commands \cite{chen2025Memory}. Additionally, Quantization-aware Training (QAT) simulates the quantization process during fine-tuning to reduce the model's precision without significantly affecting its performance.  For instance, the Edge-LLM framework adopts QAT to enable efficient LLM adaptation on edge devices \cite{yu2024edge}. The Edge-LLM framework also incorporates model pruning.  By analyzing the sensitivity of different layers of the LLM to pruning, it dynamically allocates pruning sparsity for each layer. Then, it can effectively reduce model redundancy and improve computational efficiency.

In multi-LLM edge computing scenarios,  LLM fine-tuning in this context has its unique aspects:

\begin{itemize}
    \item \textbf{Federated Fine-Tuning:} Similar to federated learning, federated fine-tuning enables multiple edge devices to collaboratively fine-tune LLMs without directly sharing raw data \cite{yu2024edge}.   Each device fine-tunes a local copy of the LLM based on its data and shares model updates with a central server or other devices.   The server aggregates these updates to improve the global model. In \cite{zhang2024towards}, Zhang et al. proposed the Federated Instruction-Tuning (FedIT) framework, as show in Fig. \ref{fig8} where the local training operations are performed at the client side, and scheduling and aggregation operations are performed at the server side.  Then, to protect data privacy, techniques like differential privacy and secure multi-party computation can be applied to the shared updates. 

\item \textbf{Split Learning for Fine-Tuning:} By segmenting the model and conducting collaborative training among multiple clients and servers, it can effectively reduce the computational and communication burden of a single client, thereby empowering the fine-tuning of the multi-LLM \cite{zhang2025split}. For instance, in the SplitLoRA framework \cite{lin2024splitlora}, only activation and gradients need to be exchanged between the client and the server, rather than the entire LLM, which significantly reduces the communication bandwidth requirements. Furthermore, in edge computing, the network conditions of different devices vary greatly. The low communication overhead feature of SL makes it more suitable for heterogeneous networks, which can better meet the deployment requirements of LLM in different regions and under different network conditions.

\item \textbf{Transfer Learning-based Fine-Tuning:} One LLM can be fine-tuned as a base model on a general edge task and then transferred to other related edge tasks for further fine-tuning \cite{dong2024fine}.  This leverages the knowledge learned from the base model, reducing the amount of data and computational resources needed for fine-tuning on new tasks.  For instance, an LLM fine-tuned on a general natural language processing task on edge devices can be transferred to specific tasks like edge-based smart customer service or smart education. It requires only minimal additional fine-tuning to achieve good performance.


\end{itemize}

\begin{figure*}[!t]
   \centering
   \includegraphics[width=7in]{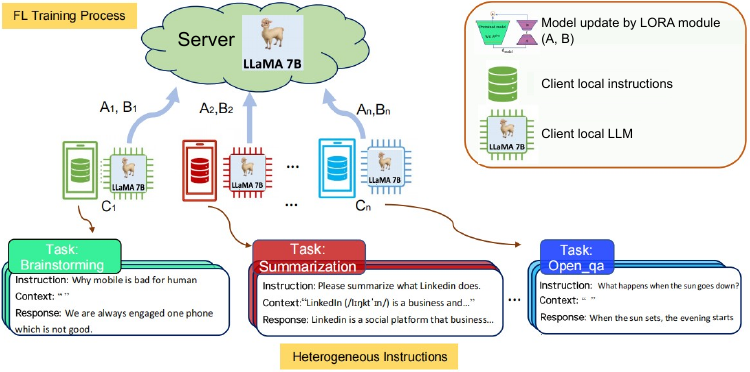}
   \caption{FedIT framework in  \cite{zhang2024towards}. All the dense layers of each LLM are introduced into a parallel LoRA module. Thus, the number of trainable parameters is significantly reduced, thereby lowering the computational and communication overhead.}
   \label{fig8}
\end{figure*}

In multi-LLM edge computing, fine-tuning faces challenges such as model compatibility, communication overhead, and data privacy protection. Ensuring the effective fine-tuning of multi-LLM while meeting the real-time and privacy requirements remains an active research area.

\subsection{Multimodal Information Fusion}

In edge computing, LLMs must process and fuse multimodal data to adapt to the complex environment of edge devices \cite{che2023multimodal}, \cite{ che2024leveraging}. In computer vision tasks like image description generation and visual question answering, LLMs can be combined with visual models. For instance, a pre-trained vision transformer extracts image features, which are then fused with text prompts input into the LLM to generate coherent and accurate textual descriptions. For instance, the LLM-Fusion model \cite{boyar2025llm} leverages LLMs to integrate representations such as selfies, text descriptions, and molecular fingerprints for precise property prediction. Furthermore, LLM can also fuse multimodal sensor data to achieve various tasks. For example, You et al.  \cite{you2025re} proposed a multimodal data fusion method based on LLM and attention mechanisms for traffic applications. It processes sensor and text data from vehicles and roads to improve traffic prediction accuracy. 

Moreover, in multi-LLM edge computing environments, multimodal information fusion exhibits different characteristics and advantages:


\begin{itemize}
    \item \textbf{Collaborative Fusion of Multiple Models:} Multi-LLM systems can collaboratively fuse multimodal data. Each LLM can process specific modalities or modal combinations. The results are then fused and cross-validated to improve fusion accuracy and robustness. For example, one LLM can focus on processing visual data while another handles textual data. The fusion results from both models are combined for comprehensive scene understanding and decision-making \cite{luo2025trustworthy}. 

    \item \textbf{Specialized Fusion Strategies:} Multi-LLM can adopt specialized fusion strategies based on the characteristics of different modalities and tasks. For instance, in multimodal recommendation systems, some LLMs can specialize in extracting user preferences from text data, while others analyze image or audio data to provide supplementary information. The fusion results help generate more personalized recommendations. A survey on multimodal recommendation systems in the LLM era highlights the advantages of integrating LLMs into multimodal recommenders, such as advanced preference summarization, context-aware fusion, and personalized content generation.

   \item \textbf{Attention-based Fusion:} This approach is highly effective in multimodal applications as it can handle noise and uncertainties in multimodal data. For example, Q-Former uses attention mechanisms to align multimodal features before generating the final output through LLM \cite{li2024multimodal}. Then, researchers directly embedded the adapter into the LLM and allowed for end-to-end training that included alignment \cite{Qwen-V2023}. In a multi-LLM system, each LLM can process different modalities or modality combinations using its attention-based fusion module. The fused results are then combined to produce the final output. This method allows for in-depth exploration of the relationships between modalities and focuses on the important information in each modality.

    \item \textbf{Cross-modal Feature Alignment:} Techniques such as the Att-Sinkhorn \cite{ma2023att} method combine the Sinkhorn metric with attention mechanisms to address the optimal transport problem between probability distributions of different modalities, thereby improving the accuracy of multimodal feature alignment. In multi-LLM systems, cross-modal feature alignment can be used to align multimodal features from different LLMs into a shared semantic space. This enables better collaboration and information sharing among LLMs, enhancing the overall performance of multimodal fusion.

\end{itemize}

Overall, multi-LLM provides unique advantages for multimodal information fusion. They are natural benefits brought by multiple perspectives.

\subsection{Lessons Learned}

According to the analysis of the above key technologies, achieving the synergy of multiple LLMs on the edge side needs to take the "performance-resource-security" triangular constraint as the core. Multimodal information fusion and model context protocols have enhanced the performance of EGI, but lightweight and resource orchestration are required to address the bottleneck of resource constraints. In addition, privacy protection and fine-tuning strategies are implemented to ensure the reliability and accuracy of the generated content. Therefore, in the future, it is necessary to focus on these three aspects to break through the triangle limitations.

\section{Trustworthy Multi-LLM for Edge Computing} \label{sec-V}

In this section, we emphasize the trustworthiness challenges of multi-LLM and design a blockchain-based approach as a tutorial. Subsequently, we also summarize the publicly available datasets related to multi-LLM, intending to contribute to the support of various EGI tasks.

\subsection{A Case Study of Trustworthy Multi-LLM for EGI}

In edge computing, devices are often in open wireless scenarios and are vulnerable to various attacks. Furthermore, edge devices are scattered, and the environment is complex, with high security threats. These factors will all weaken the response trustworthiness when multi-LLM empowers the edge. The term "trustworthiness" here primarily refers to the authenticity and reliability of the response, ensuring it can provide excellent services to users without being maliciously tampered with by attackers. Specifically, the potential factors influencing credibility in multi-LLM are:

\begin{itemize} 

\item \textbf{Difficult to Determine the Response Trustworthiness:} Different LLMs use different corpora, training methods, and scenario orientations, resulting in differences in the output for the same problem. Although there are ways of collaboration, competition, and integration, it is difficult to determine which output has the best specific credibility and quality.

\item \textbf{Threat of Malicious Behavior:} Once the deployed device is dishonest, the content generated by the LLM may generate misleading responses due to viruses, Trojans, or the operator's intentions. Furthermore, there are currently malicious models deliberately designed to actively deceive users and obtain illegal data privacy and economic benefits, such as WormGPT \cite{firdhous2023wormgpt}.

\item \textbf{Defects of the Collaboration Framework:}
Traditional Multi-LLM relies on the coordinated response of the central node, while it may be maliciously attacked or fail, thereby leading to the risk of Single-Point Failure (SPF) \cite{luo2024energy}, \cite{luo2023uls}. Meanwhile, the centralized coordination approach also has bottlenecks in terms of efficiency \cite{chen2024drdst}. Furthermore, the existing collaboration frameworks are unable to verify the credibility of LLM devices and responses, and malicious LLMs may contaminate the collaboration results \cite{luo2025convergence}.

\item \textbf{Lack of Transparency and Traceability:}
The interactive cooperation of multi-LLM, although it improves the response quality, leads to an untraceable source of the response due to the opacity of the collaboration mode. If the source of the generated content cannot be traced, it will not be easy to audit the reliability of the response \cite{liu2024blockchain}, \cite{liu2024prosecutor}.

\end{itemize}

\begin{figure}[!t]
   \centering
   \includegraphics[width=3in]{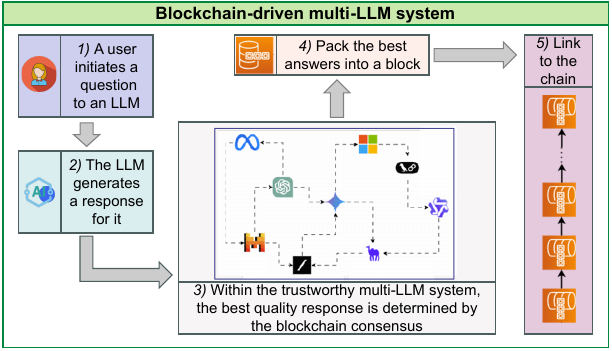}
   \caption{The blockchain-driven trustworthy multi-LLM system in \cite{luo2025trustworthy}. In a multi-LLM framework, every response generated by a specific LLM for a user undergoes verification and cross-comparison across all participating LLMs. This mechanism guarantees that the system delivers to users the most reliable answers.}
   \label{fig7}
\end{figure}

\textbf{\emph{1) Blockchain-driven multi-LLM: }} Fortunately, as a decentralized ledger technology, blockchain can drive this multi-LLM system safely and efficiently, thereby providing a reliable optimization method for edge networks. On the one hand, blockchain consensus enables multi-LLMs to respond with the best quality without relying on trusted third parties, overcoming the SPF and efficiency bottlenecks of centralized coordination \cite{lai2025accelerating},  \cite{jiang2024blockchain}, \cite{luo2025wireless}. On the other hand, the immutability and traceability of blockchain ensure the credibility of the responses generated in the multi-LLM system \cite{luo2025convergence}, \cite{luo2024symbiotic}, \cite{mao2025blockchain}. Here, based on the work in \cite{luo2025trustworthy}, we introduce how blockchain drives multi-LLM collaboration. As shown in Fig. \ref{fig7}, the specific five steps of this method are as follows:

\begin{itemize} 

\item \textbf{User Initiation:} An individual submits a query to a reliable multi-LLM system.

\item \textbf{LLM Response Creation:} Every LLM within the system formulates an answer tailored to the user's query. Subsequently, these LLMs communicate with one another through broadcast protocols within the blockchain's P2P network.

\item \textbf{Blockchain Consensus Mechanism:} Achieving consensus is crucial for identifying the best response among those generated by different LLMs. It implements a voting-based consensus to evaluate and choose the final output. The designated consensus nodes then transmit this result back to the user.

\item \textbf{Block Formation:} The top-ranked response decided by the blockchain consensus process is encapsulated into a block. To safeguard the security, immutability, and traceability of the consensus outcome, the block incorporates the hash value of the optimal solution along with its corresponding timestamp.

\item \textbf{Blockchain Extension:} Blocks holding the consensus results are appended to the blockchain and are stored in a decentralized fashion across smart devices that host the LLMs.

\end{itemize}

Furthermore, Luo et al. \cite{luo2025weighted} designed a Weighted Byzantine Fault Tolerance (WBFT) consensus based on response quality and trust value. In this consensus, the voting rights of each LLM are jointly determined by its generated content ability and credibility. The proportions of these two in the voting weights are $\alpha$ and $\beta$ respectively. This setting significantly enhances the voting rights of LLMs with high generation capabilities and credibility, and weakens the influence of malicious LLMs on the trustworthy multi-LLM system. The LLMs in this WBFT-driven trustworthy multi-LLM system are interconnected by Python mobilizing their interfaces. It runs on a high-performance server equipped with a 96-core Intel(R) Xeon(R) Gold 5220R CPU @ 2.20 GHz and 1 TB memory.

\begin{figure}[!t]
   \centering
   \includegraphics[width=3in]{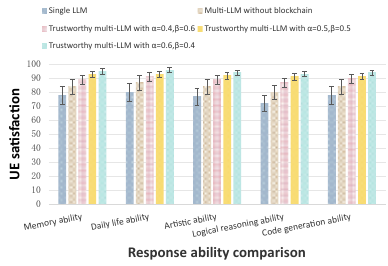}
   \caption{The comparison of users' ratings of various LLM schemes in \cite{luo2025weighted}. This result demonstrates the advantages of the multi-LLM system over a single LLM, and also reflects the importance of trustworthiness to the multi-LLM.}
   \label{fig8}
\end{figure}

Then, the authors gathered 15 volunteers from all over the world to rate the generation capabilities of individual LLMs, multi-LLM without blockchain participation, and the trustworthy multi-LLM system driven by WBFT. When the values of $\alpha$ and $\beta$ are different, the average score statistics of these schemes are shown in Fig. \ref{fig8}. This comparison result fully demonstrates the advantages of trustworthy multi-LLM driven by blockchain consensus in providing services for users, which is significantly superior to a single LLM and the multi-LLM system without blockchain participation. Also, the verification results of different values of $\alpha$ and $\beta$ reflect the importance of LLM response quality for user satisfaction. Conversely, this work also reveals the crucial impact of LLM credibility on the WBFT consensus performance.

\textbf{\emph{2) Trustworthy Multi-LLM for EGI: }}

The multi-LLM system integrating blockchain can bring more benefits to the EGI scenarios. We take the four scenarios in Section \ref{sec-III} as examples to elaborate on them respectively.

\begin{itemize}
    \item \textbf{Elderly Care:} The collaborative verification of medical advice by multiple LLMs can avoid incorrect decisions caused by the illusion or bias of a single model. On this basis, the immutability of blockchain ensures that sensitive health data is only recorded after being encrypted, meeting the compliance requirements for medical privacy. In addition, this system can also prevent the compromised wearable device LLM from sending fake alerts, such as false fall detection, ensuring the authenticity of emergency responses.
    \item \textbf{Smart Grid Inspection:}  Multiple LLMs collaboratively analyze sensor data and unmanned aerial vehicle images, and avoid missed detection by a single model through cross-validation. The participation in blockchain consensus requires the majority of LLMs to reach an agreement before triggering an alert, significantly reducing the rate of false alerts. Meanwhile, all inspection operations, such as equipment status updates and maintenance instructions, will be recorded on the chain. This will facilitate the tracking of responsible parties and prevent internal personnel or external hackers from tampering with the data.
    \item \textbf{Intelligent Transportation:} When intersection LLMs collaborate with vehicle LLMS to optimize traffic lights and path planning, blockchain consensus can prevent malicious nodes from forging traffic data, such as false congestion reports, and prevent traffic paralysis or accidents. Furthermore, vehicle sensor data needs to be verified by multiple LLMS before being used for global scheduling to prevent invaded vehicles from sending misleading information.
    \item \textbf{LAENets:} When multiple LLMs negotiate flight paths in real time, blockchain consensus can prevent malicious drones from sending false position data and causing collisions. Meanwhile, the instruction sources and execution results of logistics or communication tasks are recorded on the chain, facilitating dispute traceability and efficiency optimization.
\end{itemize}

\subsection{Open-Source Datasets}

\begin{table*}[!t]
\centering
\caption{Open-Source Dataset Related to Multi-LLM}
\label{tab:data}
\renewcommand{\arrayrulewidth}{0.8pt} 
    \renewcommand{\tabcolsep}{10pt} 
    {\fontsize{8}{10}\selectfont 
\begin{tabular}{p{1.8cm}| p{7.7cm}| p{6.3cm} }
  \hline
\textbf{Dataset} & \textbf{Description} & \textbf{Composition}  \\
  \hline

FSPO & A dataset for personalizing different LLMs & Contain a dataset with preferences \\
  \hline
BabbleBeaver & A prompt word dataset for multiple LLMs & Provide 50 typical prompt words \\
  \hline
MMLU & A dataset for testing the performance of different LLMs & Include 15,908 questions from 57 disciplines \\
  \hline
LLM-QA & A dataset for testing the selection accuracy of multiple LLMs & Contain 30 questions and answers \\
  \hline
  
GSM8K & A primary school mathematics dataset for evaluating multiple LLMs & Contain 8.5 thousand questions and answers    \\
  \hline
ChatGLM &A historically relevant dataset for pre-training multiple LLMs & Provide some prompts in JSON format   \\
  \hline
  
\end{tabular}}
\end{table*}

High-quality data plays an important role in the training and performance evaluation of multi-LLM. In this section, we outline the publicly available datasets related to multi-LLM. Table \ref{tab:data} lists the brief introductions of these datasets.

\begin{itemize}
    \item \textbf{FSPO (Few-Shot Preference Optimization):} Singh et al. \cite{singh2025fspo} designed  FSPO\footnote{https://github.com/Asap7772/fewshot-preference-optimization}. It uses a very small amount of real user preference to optimize the LLMs and make its output more in line with the preferences of specific individual users. 

    \item \textbf{BabbleBeaver\footnote{https://github.com/open-build/BabbleBeaver/tree/main}:} It provides some prompt examples for multiple LLMs, which can facilitate conversations among LLMs, including OpenAI, Google Gemini, Mistral, Anthropic, Cohere, Ollama, etc. 
    \item \textbf{MMLU\footnote{https://huggingface.co/datasets/cais/mmlu}:} This is a large-scale multi-task test dataset used to measure the model's ability to understand multi-domain knowledge and solve problems, aiming to assess the model's ability to understand and solve problems in a wide range of knowledge domains.
    \item \textbf{LLM-QA\footnote{https://github.com/collectioncard/LLM-QA-Analysis/tree/main}:} It is used to evaluate the accuracy of multiple LLMs on multiple-choice tasks. By comparing their performances, identify the model that performs best on a specific task.
    \item \textbf{GSM8K\footnote{https://openai.com/index/solving-math-word-problems/}:} It is used for training and evaluating the abilities of different LLMs in solving primary-school mathematics. Specifically, it helps the model learn to identify its own mistakes and try repeatedly until the correct solution is found. 
    
    \item \textbf{ChatGLM\footnote{https://github.com/SchweitzerGAO/awesome-chinese-chatbot}:} The dataset mentioned by ChatGLM is a history-related dataset used for fine-tuning LLMs to build a Chinese chatbot. It provides some prompts in JSON format.
\end{itemize}

\section{Future Research Directions} \label{sec-VI}

Although the capabilities of multi-LLM in model compression, resource orchestration, model context protocol, privacy protection, LLM fine-tuning and multimodal information fusion are impressive, its application in edge computing is still in its early stages. This section aims to explore the research directions related to multi-LLM-enabled EGI.

\subsection{Lightweight Multimodal LLM}

Current research has confirmed that lightweight multimodal models are of great significance for resource-constrained environments such as edge computing. For example, in the field of multimodal emotion recognition, existing studies have proposed a lightweight neural network architecture. It only uses approximately 2.7 M of parameters when analyzing multimodal information \cite{radoi2024uncertainty}. In the future, on this basis, model compression and optimization techniques such as quantization, pruning and knowledge distillation can be further explored to further reduce the model size and computational complexity, enabling it to run more efficiently on edge devices.

In terms of efficient architecture design, the design concepts of lightweight networks such as MobileNet \cite{sinha2019thin} can be drawn upon to design structures such as depth-separable convolution, suitable for multimodal data processing. This method can reduce parameter redundancy and improve computational efficiency. In addition, it is necessary to study the hardware adaptation and acceleration technologies of multimodal models in view of the characteristics of edge computing. For instance, collaborate with chip manufacturers to optimize the execution efficiency of models on specific hardware platforms, or develop specialized edge AI chips to support the efficient operation of lightweight multimodal models.

\subsection{Cross-Domain Generalization}

Realizing cross-domain generalization capability is one of the key challenges faced by multi-LLM systems. Existing studies have begun to focus on how to improve the adaptability of the model in different fields. For example, through domain adaptation algorithms such as adversarial training \cite{meegahapola2024m3bat}, the feature differences between the source domain and the target domain can be reduced, and the generalization performance of the model in new domains can be improved. In the future, it is necessary to further develop more effective domain adaptation algorithms and explore how to achieve cross-domain knowledge transfer and integration in multi-LLM systems. By constructing knowledge graphs and other means to integrate semantic information from different fields \cite{sun2024large}, it provides the model with richer background knowledge and helps it make more accurate decisions in cross-domain tasks.

In addition, optimizing the pre-training and fine-tuning strategies is also an important direction for improving the cross-domain generalization ability \cite{zhao2025movable}. Multi-domain data can be adopted for pre-training, enabling the model to learn broader general knowledge at the initial stage. In the fine-tuning stage, the method of transfer learning is adopted to make targeted adjustments to the model according to the characteristics of the target domain.

\subsection{Trustworthy AI Governance}

Improving the credibility of the multi-LLM system is the key to achieving its wide application \cite{anthuvan2025ai}. In terms of model transparency and interpretability, it is necessary to study in the future how to provide a reasonable explanatory basis for the output of the model. For example, develop an interpretation generator and adopt technologies such as attention mechanism visualization to demonstrate the concerns and reasoning paths of the model when dealing with multimodal data.

Enhanced security and robustness are also important components of trusted AI governance. The security design of the multi-LLM system needs to be further strengthened, and techniques such as adversarial training and data augmentation should be adopted to improve the robustness of the model. Meanwhile, study the vulnerability detection and repair methods of the model to promptly identify and solve potential security issues. In terms of privacy protection mechanisms, advanced cryptographic technologies are adopted to achieve collaborative processing of multimodal data without disclosing data privacy, preventing data leakage and abuse \cite{zhang2025covert}.

\subsection{Multi-LLM Reasoning}

Reasoning methods are crucial for the multi-LLM system because they significantly enhance its ability to handle complex tasks and improve the reliability of decision-making.
The multi-LLM system combined with Retrieval Enhanced Generation (RAG) and Retrieval Enhanced Perception (RAP) can handle multimodal data more efficiently and improve the accuracy and relevance of the generated content \cite{wang2025retrieval}, \cite{zhang2024interactive}. For example, when dealing with complex image description tasks, the multi-LLM system can generate more detailed descriptions by retrieving relevant image and text information. In the future, the collaborative mechanism of multiple LLMs in RAG and RAP can be explored to optimize the retrieval strategy and enhance the complementarity among models.

Additionally, Chain of Thought (CoT) can enhance the reasoning ability in complex tasks by simulating the human step-by-step reasoning process in multi-LLM systems \cite{wang2025chain}. For example, multi-LLM systems can achieve more detailed reasoning steps by decomposing problems and assigning them to different expert models. Meanwhile, the world model, as a structured framework integrating the laws and physical rules of the real world, can eliminate cognitive conflicts among LLMs and bring core advantages such as cognitive consistency and task collaboration to the multi-LLM system \cite{zhao2025world}. Furthermore, agentic AI can simulate the autonomous decision-making and interaction of agents in multi-LLM systems \cite{liu2025wireless}, \cite{zhang2024optimizing}, achieving more efficient information sharing and collaborative work to handle complex dynamic tasks.

\section{Conclusion} \label{sec-VII}

This survey comprehensively explores multi-LLM systems in edge computing, with a focus on their evolution from traditional edge artificial intelligence models to individual LLMS and then to multi-LLM systems. It discusses the typical application scenarios of this example, thereby leading to key supporting technologies such as model compression and dynamic data. Meanwhile, the survey emphasizes that multi-LLM needs to make robust decisions and provide reliable responses in an environment with high reliability and privacy. After providing relevant open-source data, the survey also discussed future research directions, including lightweight architecture, trusted governance and other issues. Overall, multi-LLM systems have demonstrated great potential in promoting the development of edge computing towards more intelligent and autonomous applications, effectively meeting the demands of our complex digital world.

\bibliographystyle{IEEEtran}
\bibliography{IEEEabrv,mylib}

\newpage


\vspace{11pt}

\vfill

\end{document}